\documentclass[useAMS,usenatbib,fleqn]{mn2e}

\usepackage{times}

\usepackage{graphicx}
\usepackage{epsfig}
\usepackage{amssymb,amsmath}
\usepackage{aas_macros}

\title[Steep power law spectra and HFQPOs]{A model of the steep power law spectra and high-frequency quasi-periodic oscillations in luminous black hole X-ray binaries}
\author[Dexter \& Blaes]{Jason Dexter$^{1}$\thanks{E-mail: 
jdexter@berkeley.edu} and Omer Blaes$^{2}$\\
$^{1}$Departments of Physics and Astronomy, University of California, Berkeley, CA 94720-3411, USA\\
$^{2}$Department of Physics, University of California, Santa Barbara, CA 93106, USA
}
\begin{document}
\pagerange{\pageref{firstpage}--\pageref{lastpage}} \pubyear{2013}
\maketitle

\label{firstpage}

\begin{abstract}
We propose a new model of the steep power law state of luminous black hole X-ray binaries.  The model uses the fact that at high luminosities, the inner radii of radiation pressure dominated accretion discs are expected to i) become effectively optically thin and ii) produce significant luminosities. The gas temperature therefore rises sharply inwards, producing local saturated Compton spectra with rapidly increasing peak energy.  These spectra sum together to form a steep power law tail to the spectrum.  A given photon energy on this tail corresponds to a narrow range in radius, so that local vertical oscillations of the disc naturally produce high quality high-frequency quasi-periodic oscillations (HFQPOs) in the hard X-ray band. The two lowest order modes have a robust frequency ratio of $\sqrt{7/3}\simeq1.53$.  This model explains the appearance of steep power law spectra and HFQPOs at high luminosity, the 3:2 HFQPO frequency ratios, and their association with the power law spectral component. We predict an increase in QPO quality factor when the power spectrum is restricted to a narrower photon energy band, and an increase in HFQPO frequency at higher X-ray energies or lower luminosities. Future X-ray telescopes could detect additional HFQPOs from higher order modes. We demonstrate how this model could be used to measure black hole spin from HFQPOs, and qualitatively estimate the spin of GRO J1655-40 as $a/M\sim 0.4 - 0.7$.
\end{abstract}


\section{Introduction}

Intensive monitoring of black hole X-ray binaries (BHBs) has established a set of characteristic outburst states classified by the X-ray luminosity, colour (``hardness''), rms variability, and spectral shape (e.g. \citealt{tanakashibazaki1996,remillardmcclintock2006}, hereafter RM06). At the highest observed luminosities, the spectrum is dominated by a steep (photon index $\Gamma > 2.4$) power law extending in some cases to MeV energies \citep{groveetal1998}. This ``steep power law'' or ``high intermediate'' state is associated with quasi-periodic oscillations (QPOs). The high-frequency ($> 50 \rm Hz$) QPOs (HFQPOs) are of particular interest, as their frequencies are comparable to the expected orbital frequency near the innermost stable circular orbit (ISCO) of a stellar mass black hole. HFQPOs have been observed in seven BHBs, with typical amplitudes $\sim 0.01$. In some cases twin frequencies appear with a 3:2 ratio \citep[e.g.,][]{strohmayer2001}. They are associated with the power law (hard) rather than the accretion disc (soft) spectral component (RM06).  Observations of HFQPOs are a promising means to measure black hole spin \citep{abramowiczkluzniak2001} and probe strong gravity, provided that a model for their origin can be found.

Many models have been proposed for HFQPOs, e.g. oscillations in accretion discs \citep[][]{stellaetal1999,wagoner1999,kato2001} or orbiting hotspots \citep{kato2004,schnittmanbertschinger2004}, but from these models it is difficult to determine why a 3:2 frequency ratio should be seen, why the QPOs should be associated with the power law spectral component, and why they should appear only at high luminosity. Numerical MHD simulations of black hole accretion discs do not find significant power from orbiting hotspots \citep{schnittman2006,dexter2009}.  On the other hand, localized vertical oscillation modes are observed in vertically stratified radiation MHD shearing box simulations \citep{hirosestable2009,blaesetal2011}.  The two strongest modes are the vertical epicyclic oscillation and lowest-order acoustic breathing mode. In a radiation-dominated medium (with adiabatic index $\gamma=4/3$), the period ratio of these modes is $\sqrt{1+\gamma} \simeq 1.53$, nearly identical to the observed 3:2 ratio of HFQPOs in several sources. This ratio is completely independent of the detailed vertical structure \citep{blaesetal2011}. These vertical oscillations have also been observed in global MHD simulations \citep{reynoldsmiller2009}, where they appear to be independently excited at each radius, so that their frequencies scale approximately as $\sim r^{-3/2}$. The modes will therefore blend together into broadband noise when averaging the emission over the disc, unless some mechanism exists which can filter the power from a narrow range in radius. 

The origin of the steep power law spectrum observed in BHBs at near Eddington luminosities is also poorly understood. A hot, tenuous corona could Compton scatter the bulk of the disc seed photons into a power law \citep{donekubota2006}. Radiative transfer calculations from an MHD simulation have shown that steep power law spectra can be produced in this way if enough of the accreting material goes into the corona rather than the disc \citep{schnittmanetal2013}. However, this scenario does not naturally explain the origin or properties of observed HFQPOs.

We propose an alternative scenario for the steep power law state, where the accretion disc itself produces the entire X-ray spectrum. We   use a semi-analytic accretion disc model to calculate approximate spectra (\S\ref{sec:spectral-calculation}) and HFQPO profiles (\S\ref{sec:filt-high-freq}), demonstrating that high quality oscillations are naturally associated with steep power law spectra, and that both only appear at high luminosities. We give a qualitative example of measuring spin from HFQPO and spectral observations of GRO J1655-40 in \S \ref{sec:parameter-space-spin}, and discuss further implications and predictions of the model in \S \ref{sec:discussion}.


\section{Steep power law spectra and quasi-periodic oscillations from an accretion disc}
\label{sec:sample-disc-model}

\begin{figure}
\begin{center}
\includegraphics[scale=0.7]{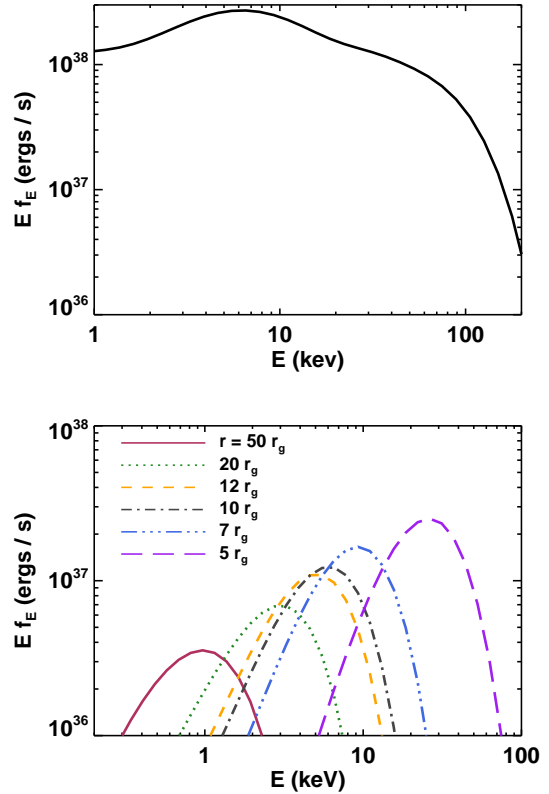}
\caption{\label{spectra}Sample disc-integrated spectrum (top) and contributions from several annuli (bottom). Many radii contribute to the spectral peak, while the rapid increase in gas temperature in the effectively optically thin annuli lead to well separated spectral peaks, whose  sum leads to a steep power law tail. The local spectra are saturated Compton (Wien) spectra, whose luminosities are significantly suppressed in comparison to blackbody spectra at the same temperature. The parameters are $a/M = 0.5$, $\dot{m}=0.7$, and $\Delta \epsilon = 0.1$. The resulting luminosity is $L/L_{\rm edd} \simeq 1.0$.}
\end{center}
\end{figure}

\begin{figure}
\begin{center}
\includegraphics[scale=0.7]{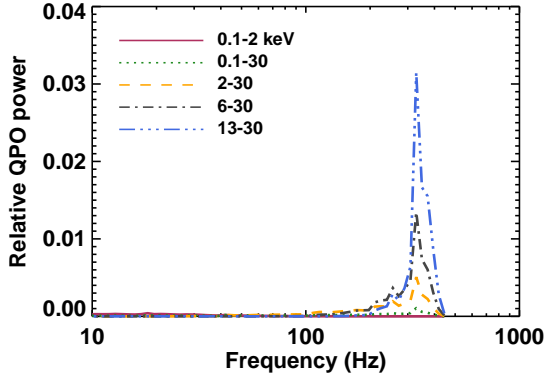}
\caption{\label{qpo}Synthetic QPO profiles from our sample disc model. Annuli spectra are assumed to contribute to the power at the local vertical epicyclic frequency, and are integrated over various bands in energy and then normalized. At low energies, many radii contribute, resulting in a broad flat profile. A sharp QPO-like peak is produced at higher energies where a narrow range in radius dominates the total flux. The quality factor, measured as $\nu / \rm FWHM$ for a Gaussian profile, is $Q = 2.1$, $3.8$, $4.9$ for energy bands of $2-30$, $6-30$, and $13-30$ keV. The model parameters are $a/M = 0.5$, $\dot{m}=0.7$, and $\Delta \epsilon = 0.1$. Compare to the observed $300$ Hz QPO in GRO 1655-40 \citep{remillardetal1999}.}
\end{center}
\end{figure}

Accretion disc spectra at luminosities comparable to Eddington could naturally provide a filter for HFQPOs, as long as sufficient radiation can be generated by the inner radii. At high luminosities, radiation-dominated discs are expected to become effectively optically thin. Global, general relativistic MHD simulations have found that the inner radii can still produce a significant amount of emission \citep[e.g.,][]{noble2009}. \citet{zhudavisetal2012} showed that the combination of these factors can lead to a rapid increase in gas temperature in the inner radii, leading to a high energy power law spectral tail from the superposition of saturated Wien spectra. At high luminosity, this power law tail becomes dominant, consistent with the observed steep power law spectra in BHBs where the HFQPOs are seen (figure \ref{spectra}). The gas temperature, and the peak energy of the resulting Wien spectra, rise rapidly inwards. Therefore, the power law tail is naturally associated with a narrow range of radius. This effect can explain the presence of sharp HFQPOs ($\nu / \rm FWHM > 2$) in steep power law spectra and their association with high energy photons (figure \ref{qpo}).

We demonstrate the plausibility of this scenario for the steep power law state by estimating spectra and QPO power profiles from the semi-analytic thin accretion disc model of \citet[][AK00]{agolkrolik2000}. This model is a  generalization of the standard thin disc model \citep{shaksun1973,novthorne} allowing for a non-zero stress and luminosity at the ISCO. Effective temperature profiles from the AK00 model are in reasonable agreement with current GRMHD simulations \citep[][]{noble2009}. The AK00 model only adds a single parameter to standard accretion disc models, $\Delta \epsilon$, the increase in overall radiative efficiency relative to a standard accretion disc, and is therefore a simple way to capture the required physics. However, we note that a more complicated slim disc model \citep[e.g.,][]{abramowiczetal1988} would provide a more realistic treatment of accretion physics at high luminosity, where advection is important and the disc is no longer thin. The other model parameters are the \citet{shaksun1973} viscosity parameter $\alpha$, the normalized accretion rate ($\dot{m} \equiv \epsilon \dot{M} c^2 / L_{\rm edd}$), and the black hole mass ($M$) and spin ($a/M$). Here we fix $\alpha = 0.02$ and $M = 7 M_\odot$ and vary the other parameters. 

\subsection{Spectral calculation}
\label{sec:spectral-calculation}

We use the AK00 surface density, effective temperature, and scale height profiles to calculate the gas temperature and spectrum from each disc annulus, spaced evenly in $\log r$. We use a one zone model (ignoring vertical structure), and assume that radiation pressure dominates throughout the disc. The local spectrum is calculated using the approximate solutions for Comptonised bremsstrahlung with $y \gg 1$ described in \citet{ryblight} \S7.6. The temperature and spectrum are calculated iteratively by matching the integrated local spectrum to the flux required by the model at that radius: $F(r) = \int d\nu I_\nu (r)$, where $I_\nu$ is the intensity. (We neglect relativistic factors in this spectral calculation, though they are included in the disc structure model.)  

\begin{figure}
\begin{center}
\includegraphics[scale=0.6]{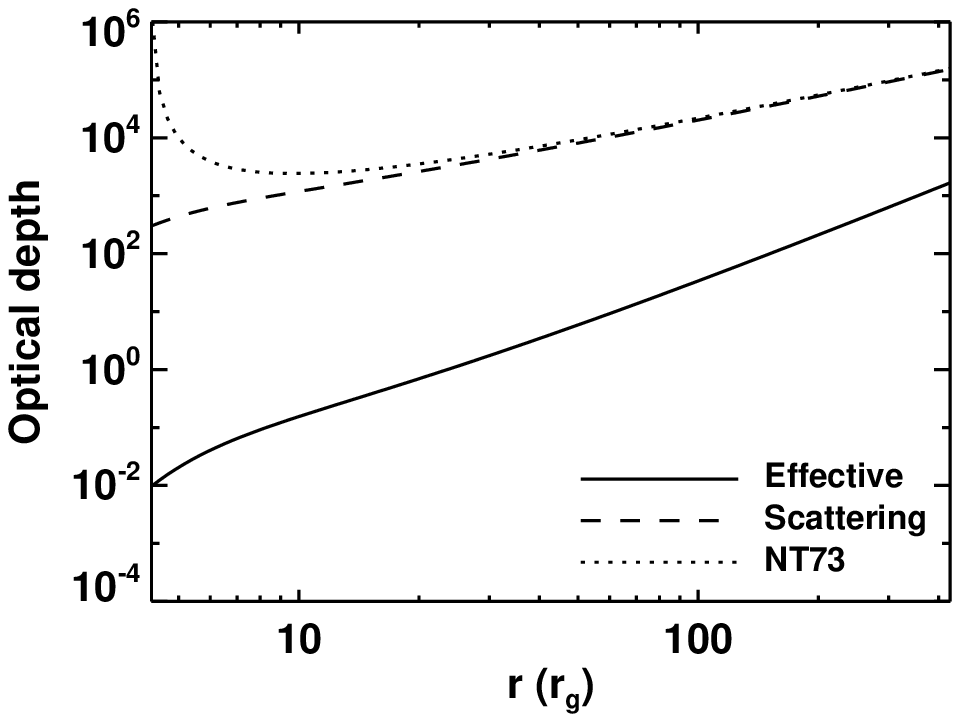}\\
\includegraphics[scale=0.6]{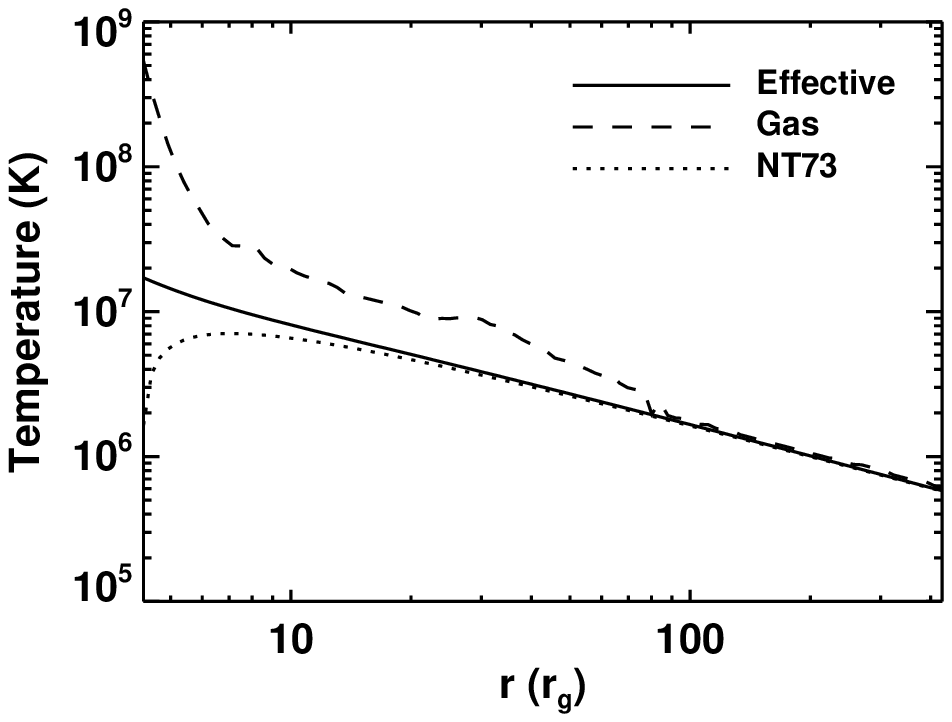}
\caption{\label{tautemp}Measures of optical depth (top) and temperature (bottom) as functions of radius for the model in figures \ref{spectra} and \ref{qpo}. The gas temperature becomes significantly larger than the effective temperature in the inner radii where the absorption opacity is small.}
\end{center}
\end{figure}

In optically thick annuli at relatively large radius, the intensity is a modified blackbody spectrum with a colour-correction factor $\lesssim 2$. Effectively optically thin annuli instead produce a Wien spectrum with a total flux of $I_\nu = H A(\rho,T) \epsilon_\nu (\rho, T)$, where $\epsilon_\nu$ is the bremsstrahlung emissivity, $H$ is the disc scale height, and $A$ is the amplification factor of the local bremsstrahlung emission by inverse Compton scattering. The Wien intensity is much smaller than the blackbody intensity at the same temperature in the inner radii where absorption is inefficient. In order to  maintain the required effective temperature, the gas must become extremely hot, reaching gas temperatures a factor $\lesssim 20$ larger than the effective temperature near the ISCO for $L \sim L_{\rm Edd}$ (figure \ref{tautemp}). This does not occur in NT73 discs, where the effective temperature artificially drops to zero at the ISCO.

Sample spectra from the full disc and several local area-weighted annuli are shown in figure \ref{spectra}. The rapid increase in gas temperature in the inner radii leads to a superposition of Wien spectra with higher and higher energy peaks. Although their normalizations also increase moving inwards, fewer annuli contribute to the emission, creating a power law tail with a photon index of $\Gamma \simeq 2-3$, in agreement with observed steep power law spectra (RM06). On the steep power law tail portion of the spectrum, the flux at a given energy is produced by a narrow range in radius.

\subsection{A filter for high-frequency quasi-periodic oscillations}
\label{sec:filt-high-freq}

We roughly estimate the HFQPO component of the power spectrum of our model by assuming that the local QPO flux amplitude is proportional to the total local X-ray flux. A linear rms-flux relation is seen locally in radiation MHD simulations and is also seen observationally \citep{uttleymchardy2001}. We further assume that the QPO amplitude is proportional to the rms noise.  

Using this approximation, the QPO profile is calculated by integrating the flux in each annulus over a desired energy band, and dividing by the total flux in that energy band. Each annulus contributes to the QPO amplitude at a different frequency, which we take to be the vertical epicyclic frequency at that radius for a chosen black hole mass and spin. The profile is then normalized by the total flux in the band, and squared to convert to power. A sample QPO profile is shown in figure \ref{qpo} corresponding to the spectrum in figure \ref{spectra}. The parameters are chosen to reproduce the observed $2 \nu_0$ oscillation in GRO 1655-40 at $\nu \sim 300$ Hz (cf. RM06 figure 11). When the flux of a chosen energy band is dominated by optically thick annuli, a wide range of radii contribute and the profile is flat with no QPO (e.g., $0.1-2$ keV). When the flux in the energy band is dominated by optically thin annuli, QPOs naturally emerge from the narrow range in radius corresponding to the band. The QPO profiles are narrower (higher $Q \equiv \nu / \rm FWHM$) when integrated over smaller energy bands. The Q values found are $\simeq 2-6$ for hard energy bands ($> 2$ keV) depending on the model parameters and energy band, consistent with observed HFQPOs.

\begin{figure}
\begin{center}
\includegraphics[scale=0.65]{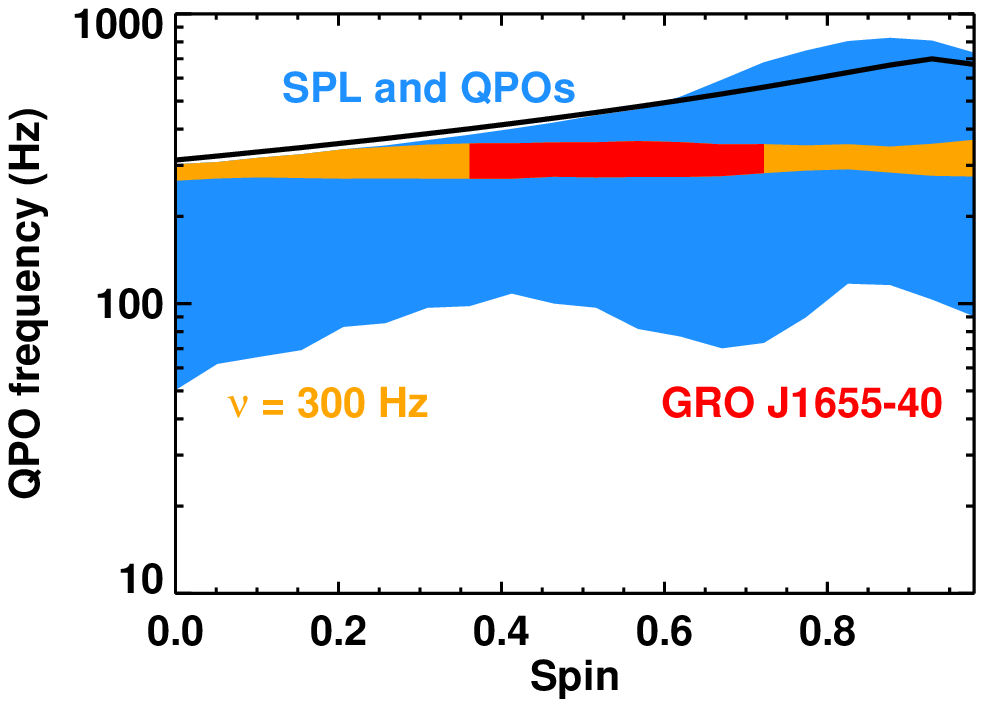}\\
\includegraphics[scale=0.65]{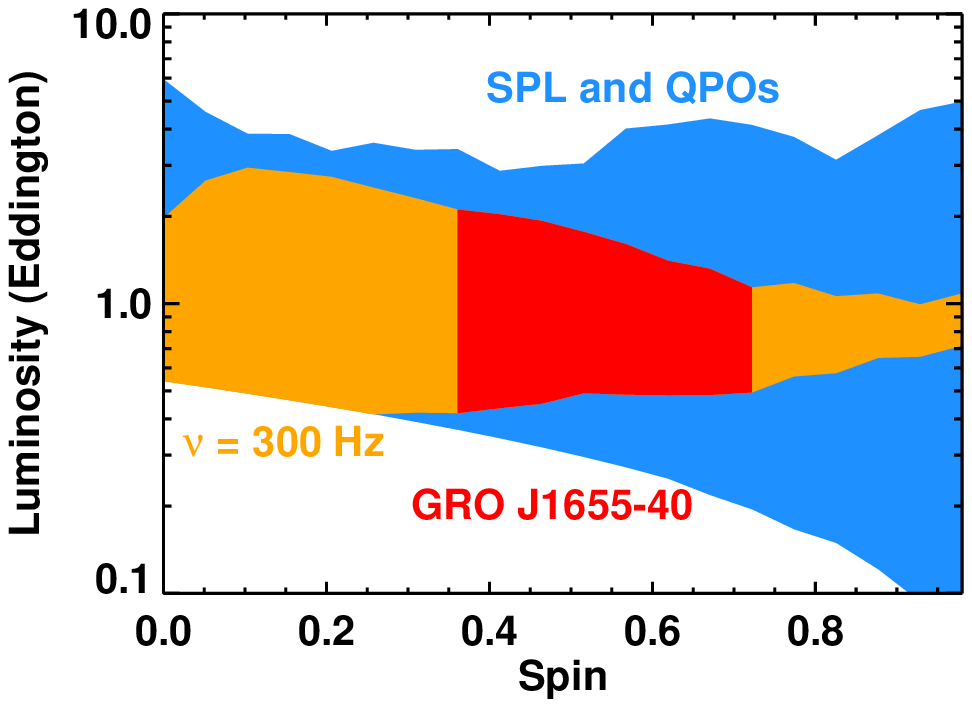}
\caption{\label{gro}Regions of model parameter space in QPO frequency (top) and luminosity (bottom) vs. black hole spin. The different shaded regions correspond to i) spectra with $\Gamma = 2.4-3.0$ and QPOs with $Q > 2$ integrated over 6-30 keV (blue), ii) the subset with QPO frequencies $\nu = 300 \pm 15\%$~Hz (orange), and iii) the subset where the QPO frequency is stable over factors $> 2$ in luminosity  and where $\Gamma = 2.4-2.6$ as seen in GRO J1655-40 (RM06, red). We use these constraints to qualitatively estimate a spin of $a/M \sim 0.4-0.7$ for GRO J1655-40.}
\end{center}
\end{figure}

\subsection{Parameter space and spin dependence}
\label{sec:parameter-space-spin}

Figure \ref{gro} shows the portion of parameter space where the disc model produces power law spectra with $\Gamma = 2.4 - 3.0$ along with QPOs ($Q > 2$). Such models are possible at all black hole spin values over a range of luminosity, producing a wide range of HFQPO frequencies at each spin. Power law spectra require higher luminosities for small black hole spin, because the effective optical depth is higher at fixed luminosity. The QPO frequency increases with spin along with the underlying vertical epicyclic mode frequency. The maximum QPO frequency at each spin is the mode frequency at the ISCO, the model inner boundary. Physically, we expect that standing vertical oscillations should not exist inside that radius, although exactly where the modes are present in radius should be investigated with future simulations.

A large range of QPO frequencies are possible at each value of black hole spin. However, it is still possible to have relatively stationary QPOs ($< 15\%$ changes) over ranges of factors $\lesssim 5$ in luminosity as observed \citep[e.g.,][]{remillardetal2002}, as long as the luminosity changes derive from variations in the accretion rate. Changing the luminosity by changing $\Delta \epsilon$ instead leads to large changes in the spectral shape and QPO frequency with luminosity, inconsistent with observations. The favored value is $\Delta \epsilon / \epsilon \simeq 1$, or equivalently a smooth effective temperature profile with radius (figure \ref{tautemp}). The radius where the QPO emerges is $r \gtrsim r_{\rm ISCO}$, and measuring black hole spin from observed HFQPOs requires measuring this radius using spectral information. We demonstrate this procedure using the X-ray binary GRO J1655-40 as an example. The measured $300 \rm Hz$ QPO constrains the parameter space, but solutions still exist at all values of black hole spin. The stability of HFQPOs with factor of few changes in luminosity limits the spin to $a/M \lesssim 0.7$, and the observed photon indices $\Gamma \simeq 2.4-2.6$ (RM06) would lead to an estimate $a/M \sim 0.4-0.7$. This is comparable to the spin measured using the thermal continuum method ($0.7\pm0.1$, \citealt{shafeeetal2006,davisdoneblaes2006,mcclintocketal2013}), but lower than that inferred from fluorescent iron line measurements ($>0.9$, \citealt{reisetal2009}) and higher than a recent estimate from the relativistic precession QPO model \citep[$\simeq 0.29$][]{mottaetal2013}. This is a qualitative example. Quantitatively, the spin constraint is subject to systematic uncertainties from our choice of disc model (surface density and effective temperature profiles) and approximations used in the spectral calculation.

\section{Discussion}
\label{sec:discussion}

We have proposed a novel model for steep power law state spectra and HFQPOs in BHBs.  We propose that HFQPOs are due to local vertical epicyclic and acoustic breathing modes, whose frequency ratio is $\simeq 1.53$ in radiation pressure dominated media, independent of vertical structure. A standard disc becomes effectively optically thin well outside the ISCO at high luminosity, and these annuli must become extremely hot to produce significant luminosities. We have shown that saturated Wien spectra from these annuli can lead to the observed steep power law spectrum, and simultaneously provide a filter to associate the observed X-ray energy band with a narrow range in radius, and hence a narrow range in HFQPO frequency. This model can therefore simultaneously explain the steep power law state spectra and HFQPOs within the context of standard thin accretion disc theory. We have given an example for how to use the model along with observed HFQPOs and spectral properties to estimate black hole spin, qualitatively inferring $a/M \sim 0.4-0.7$ for GRO J1655-40.

For simplicity, we have used the AK00 model, which is based on standard thin disc accretion theory but can explain luminosity profiles from numerical MHD simulations of black hole accretion discs \citep{shafee2008,noble2009}. However, it is not clear whether standard thin disc accretion theory can be applied when the disc is radiation-dominated. Recent radiation MHD simulations have found that the disc may be thermally unstable (Jiang, Stone \& Davis 2013, submitted), and the non-linear saturation of the instability has not yet been studied. Global simulations of the entire disc will be needed to determine its structure, but it may still be that it resembles that of a standard thin disc.  Observationally, there is little evidence for the thermal instability in BHBs, except perhaps in some sources reaching very high luminosities \citep{doneetal2004}.  The important assumptions of our model are that the surface density decreases with decreasing radius, so that at high luminosities the inner radii become effectively optically thin, and that these radii have substantial luminosities (high effective temperatures). 

We have used approximate analytic solutions for the emergent spectrum and ignored the vertical disc structure. The results qualitatively agree with more accurate radiative transfer calculations \citep{zhudavisetal2012,taoblaes2013}. The gas temperature should continue to increase inside the ISCO, and including these annuli would likely extend the spectrum to higher energies ($\gtrsim 100 \rm keV$). Relativistic effects, including Doppler beaming, are also likely to modify the spectrum further from what is assumed here, and a proper ray tracing calculation (e.g. \citealt{dexteragol2009}) of the model spectrum and HFQPOs will be necessary to make more robust predictions.  Extending the power law emission to photon energies beyond an MeV \citep{groveetal1998} may also require a nonthermal electron population. We have also ignored the effects of radiative advection \citep{abramowiczetal1988}, which become important at the high luminosities considered. \citet{donekubota2006} further showed that in some cases BHB spectra become softer at high luminosities, implying that the disc is truncated outside of the ISCO, unlike what we have assumed here. The present treatment is sufficient to demonstrate the promise of the model, and we intend to address many of these issues in the future.

The amplitudes of simulated modes relative to the noise is uncertain, but we can estimate the local mode amplitudes required to explain the observed HFQPOs using our model. For the acoustic breathing mode (proposed to be $3 \nu_0$), the flux perturbation is $\Delta F / F \sim \Delta v / c_s$, where $\Delta v$ is the velocity perturbation and $c_s$ is the sound speed. For the observed QPO amplitudes of order a per cent, a small velocity perturbation should suffice to excite the mode to observable levels. The vertical epicyclic mode is a displacement of the fluid in the vertical direction, and does not directly lead to a flux perturbation. In order to produce an observable QPO, this motion must be detectable, e.g., from Doppler beaming. The flux perturbation for Doppler beaming would be $\Delta F / F \sim v / c \sim \xi / r$, where $\xi$ is the vertical displacement. At small radius, where the scale height is large ($H/r \sim 1$), this is a small displacement. 

Observationally, the HFQPOs forming the stable 3:2 ratio pair do not always appear simultaneously. This would not be surprising if they correspond to the vertical epicyclic and acoustic breathing modes, given their distinct excitation physics. The vertical epicyclic mode requires a vertically asymmetric momentum ejection to displace the local disc centre of mass.  The observed trend that the vertical epicyclic mode is seen at higher power law luminosities \citep{remillardetal2002} is interesting since those observations may be closer in time to the observed transient jet ejections, which could excite vertical disc oscillations. 

In our model, the disc is optically thick to electron scattering at all radii ($\tau \gtrsim 100$, figure \ref{tautemp}), and the photon diffusion time out from the midplane is a factor of $\sim 10$ longer than the mode period. However, in current radiation MHD simulations the breathing mode amplitude grows linearly with height up to the photosphere, where the diffusion time becomes short, allowing the mode to produce flux variations at each radius. We have assumed that the modes are excited independently at each radius, as found in the global simulations of \citet{reynoldsmiller2009}. Recent measurements find stable phase lags over the range of QPO frequencies \citep{mendezetal2013}, perhaps suggesting at least some radial coherence for the modes.  In addition to our spectral filtering mechanism, trapping the modes in the radial direction would allow for higher quality factors and more stable QPO frequencies than found here. For example, localised rings of pressure and density, such as might form in the development of a radiation pressure driven ``viscous" instability \citep{lightmaneardley1974}, could create coherent axisymmetric oscillations \citep{blaes2006}.

The proposed model makes several predictions testable with current archival data and future X-ray missions. First, we predict that restricting the energy band used to measure the power spectrum should lead to higher quality HFQPOs, since this is effectively selecting for a narrower set of radii in the accretion disc. In addition, the measured HFQPO frequencies are not necessarily expected to match the mode frequency at the ISCO or at any particular radius. Small variations in the radius where the QPO profile peaks should lead to frequency changes of  $\gtrsim 10-20\%$ as the luminosity varies by factors of several, and shifts of similar magnitudes are reported by \citet{bellonietal2012}. We further predict that the QPO frequency should decrease with increasing luminosity, as disc temperatures corresponding to a particular photon energy band move to larger radius. Detections of HFQPOs in hard X-rays may also find shifts of the HFQPO to higher frequency. 


The vertical epicyclic mode and the lowest order acoustic breathing mode are the most pronounced in current local radiation MHD simulations of black hole accretion discs, but several other higher order modes are present as well. These could lead to additional HFQPOs at higher frequency. Although we have assumed these two modes, the proposed steep power law spectrum is a general mechanism for filtering modes from a narrow range of radii, and could apply to other models for disc oscillations or orbiting inhomogeneities. Further, if HFQPOs are the result of narrow ranges in radius producing certain energy ranges of steep power law spectra, then this should have consequences for the broadband noise as well. For example, the power spectrum slope or break frequency may depend on the chosen X-ray energy band. The B type low-frequency QPOs are correlated with HFQPOs in the steep power law state  \citep{remillardetal2002}, but are left unexplained by our model. Future studies aiming to explain the origin of these features may be guided by the framework for the steep power law spectrum and HFQPOs presented here.

\section*{acknowledgements}
We thank D. Altamirano, S. Davis, C. Done, J. Krolik, and E. Quataert  for stimulating discussions related to this work, as well as the referee, O. Straub, for useful comments. This work was partially supported by the University of California High-Performance Astrocomputing Center.

\footnotesize{
\bibliographystyle{mn2e}
\bibliography{master}
}
\label{lastpage}

\end{document}